\documentclass[%
preprint,
amsmath,amssymb,
aps,
prb,
]{revtex4-1}
\usepackage{graphicx}
\usepackage{bm}
\usepackage{multirow}
\usepackage{rotating}
\usepackage{float}
\usepackage{dcolumn}
\usepackage{amsmath}
\usepackage{amssymb}
\usepackage{enumerate}
\usepackage[shortlabels]{enumitem}

\begin{document}

\title{Interplay of electronic structure and unusual development in crystal structure of YbAuIn and Yb$_{3}$AuGe$_{2}$In$_{3}$}

\author{Zs. R\'{a}k}
\email{zrak@ncsu.edu}
\affiliation{Department of Materials Science and Engineering, North Carolina State University, Raleigh, 27695-7907}

\date{\today}

\begin{abstract}

First-principles calculations within the DFT are employed to investigate the relationship between the electronic structure and the unexpected features of the hexagonal cell parameters of YbAuIn and Yb$_{3}$AuGe$_{2}$In$_{3}$. Calculations indicate that YbAuIn is an intermediate valent system with one Yb 4\textit{f} state pinned to the Fermi level, while Yb$_{3}$AuGe$_{2}$In$_{3}$ is closer to integer valency with all Yb 4\textit{f} states occupied. Structural relaxations performed on LaAuIn and LuAuIn analogs reveal that expansion of the \textit{c}-parameter in Yb$_{3}$AuGe$_{2}$In$_{3}$ is attributable to larger size of the divalent Yb compared with intermediate valent Yb.

\end{abstract}


\maketitle 

\section{\label{intro}Introduction}

Among strongly correlated materials, Yb-containing intermetallics are particularly fascinating, because they display a variety of intriguing physical properties arising from the presence of the  localized 4\textit{f} states~\cite{Lawrence1981}. Yb exhibits two valence states: magnetic, trivalent (Yb$^{3+}$) or nonmagnetic, divalent (Yb$^{2+}$) configurations. Depending on the electronic and crystallographic environment, as well as on external conditions of temperature and pressure, the two possible valence configurations can be energetically degenerate, giving rise to mixed and intermediate valency, magnetic transitions induced by pressure or alloying, anomalous low temperature thermodynamic and  transport properties~\cite{Stewart2001, Zell1981}. Furthermore, because of the hybridization between the localized 4\textit{f} states and delocalized s and p states many Yb-based intermetallics exhibit heavy fermion behavior, Kondo coupling, Ruderman-Kittel-Kasuya-Yosida interaction~\cite{Lawrence1981, Stewart2001, Fisk1988, Stankiewicz2012, Wigger2007, Chondroudi2007, wu2008, Trovarelli2000, Gegenwart2002, Bauer1991}.  Families of isostructural Yb-containing systems display a wealth of ground state properties and different low energy excitations as a function of composition. Far from being exhaustive, examples of such systems include YbMCu$_{4}$ (M=Ag, Au, Pd, In, Cu, Cd, Mg, Tl, Zn)~\cite{Rossel1987, Severing1990, Hiraoka1995, Kojima1995, Continenza1996, Sarrao1996, Sarrao1999, Antonov2000}, YbTX (T=transition metal, X=Sn, Bi)~\cite{Kaczorowski1999, Pietri2000, Szytula2003}, and Yb$_{2}$T$_{2}$In (T=Cu, Pd, Au)~\cite{Giovannini2001}.

\begin{figure}
\centering
\includegraphics[scale=0.15]{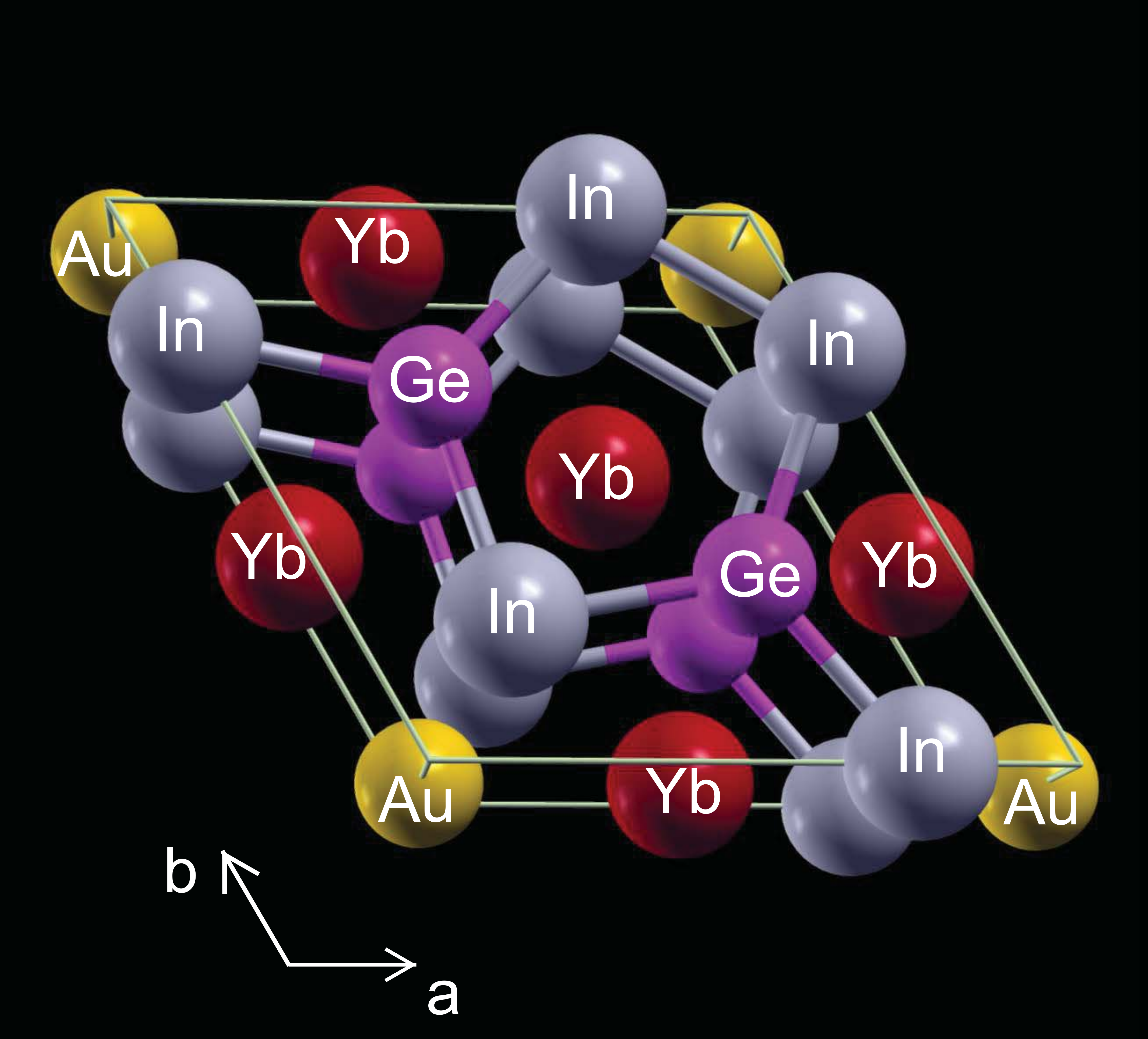}
\caption{(Color online)Hexagonal unit cell of Yb$_{3}$AuGe$_{2}$In$_{3}$. The structure cen be regarded as alternating monoatomic layers of In$_{3}$Ge$_{2}$ (or In$_{3}$Au$_{2 }$in the case of YbAuIn) and Yb$_{3}$Au.}
\label{fig1}
\end{figure}

The present theoretical investigation was motivated by a recent experimental study on the crystal structure, electronic and magnetic properties of the newly synthesized Yb$_{3}$AuGe$_{2}$In$_{3 }$ and its ternary analog YbAuIn~\cite{Chondroudi2011}. As shown in fig.~\ref{fig1}, the new compound Yb$_{3}$AuGe$_{2}$In$_{3}$ has a hexagonal crystal structure ($P\bar{6}2m$) and it is obtained by replacing two Au atoms in the structure of YbAuIn~\cite{Rossi1977} with two Ge atoms.

The crystallographic analysis of the two Yb-containing systems reveals a peculiar feature~\cite{Chondroudi2011}: when the two Au atoms in YbAuIn are replaced by Ge, the hexagonal unit cell exhibits a contraction along the \textit{a} and \textit{b} axes and an expansion along the \textit{c} axis. On the one hand, the fact that the \textit{a}- and \textit{b}-parameters of YbAuIn (\textit{a}=\textit{b}= 7.712 \AA) are larger than those of Yb$_{3}$AuGe$_{2}$In$_{3}$ (\textit{a} = \textit{b} = 7.315 \AA) makes sense because the Au atom is larger than Ge, but on the other hand, the observation that the \textit{c}-parameter of YbAuIn (\textit{c} = 4.029 \AA) is 10\% shorter than that of Yb$_{3}$AuGe$_{2}$In$_{3}$ (\textit{c} = 4.421 \AA) cannot be explained by the size difference between Au and Ge. The present work addresses the puzzle of this unusual structural change using electronic structure calculations on YbAuIn and Yb$_{3}$AuGe$_{2}$In$_{3}$, focusing on the role played by the Yb 4\textit{f}-electrons in the electronic and structural properties of the two systems.

\section{\label{method}Method of calculation}

The calculations have been carried out using the highly accurate full potential linearized augmented plane wave method (FP-LAPW)~\cite{Singh2005} within density functional theory (DFT)~\cite{Hohenberg1964, Kohn1965} as implemented in the WIEN2k program~\cite{Schwarz2002}. The exchange and correlation potential was estimated by the local spin density approximation (LSDA)~\cite{Perdew1992}, necessary for magnetic systems. To overcome the inability of DFT to model localized states, on-site Coulomb interaction was added to the Yb 4\textit{f} states using the ``around-mean field'' version (AMF) of the LSDA+\textit{U} formalism~\cite{Czyzyk1994}.  According to the literature physically reasonable values for the screened Coulomb repulsion (Hubbard \textit{U} parameter) within the Yb 4\textit{f }states are in the range of 6-7 eV. To check the dependence of the results on the particular value of the \textit{U} parameter, the calculations have been performed for \textit{U} = 5.44, 6.75, and 8.16 eV (corresponding to 0.4, 0.5, and 0.6 Ry). Based on energy convergence tests, the calculations were carried out using the following setup: the muffin-tin radii values (in atomic units; 1 au = 0.529 \AA) were chosen as 2.5 for all the atoms, the cutoff parameter was RK$_{max}$ = 9.5, and the energy of separation between valence and core states was $-6.0$ Ry. Self-consistent iterations were performed with 35 k-points in the irreducible part of the Brillouin zone (IBZ), and convergence was assumed when both the energy and the charge difference between two consecutive iterations were less than 0.0001 Ry and 0.001e respectively. Scalar relativistic corrections were added and spin-orbit coupling (SOC) was incorporated using second variational procedure~\cite{Koelling1997}. With SOC included, the energy range in which eigenvalues were searched, was set from $-9.0$ to 7.0 Ry. The calculations were carried out using the experimentally determined lattice parameters.

\section{\label{results}Results and discussions}

\begin{figure}
\includegraphics[scale=0.5]{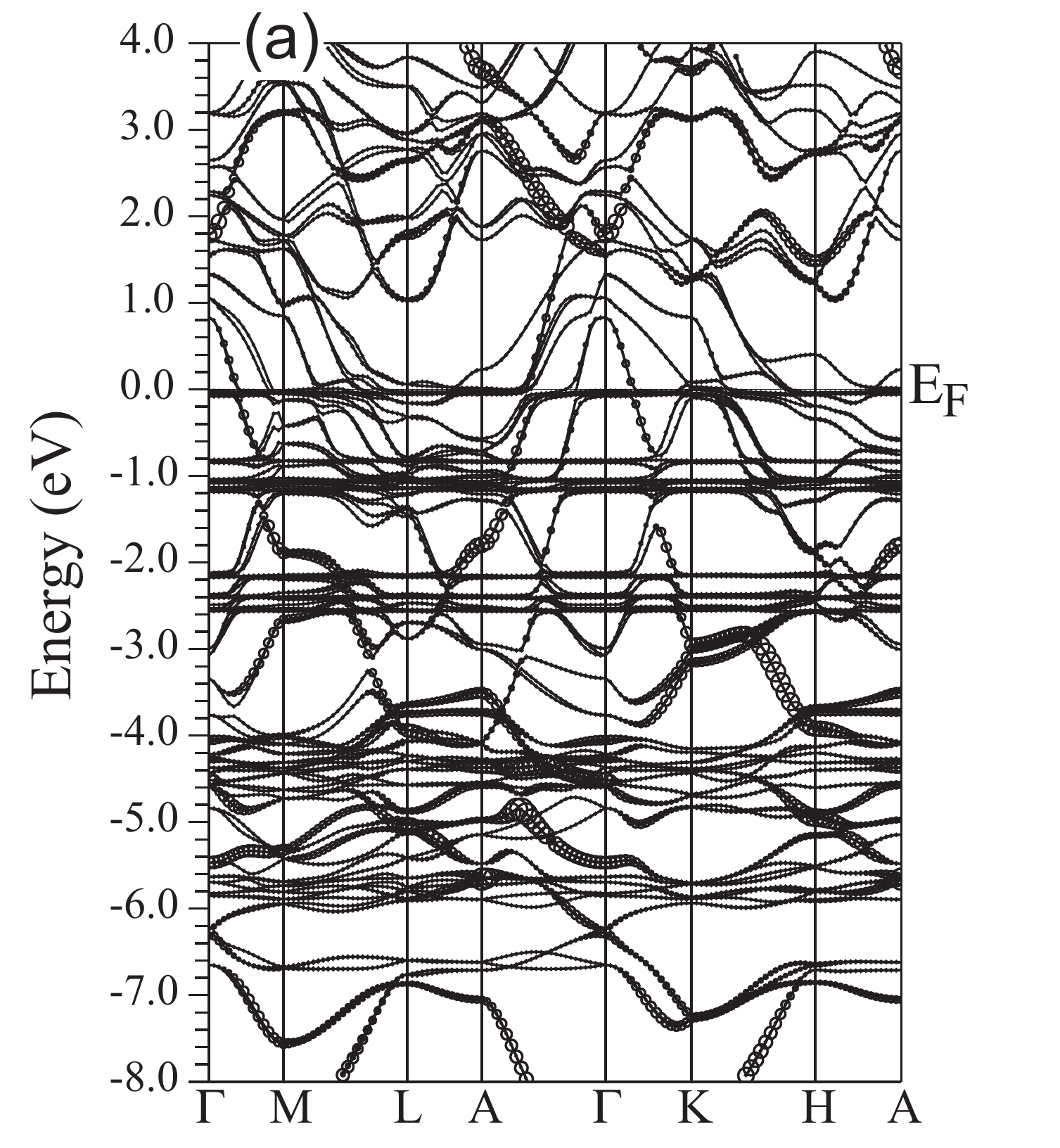}
\includegraphics[scale=0.28]{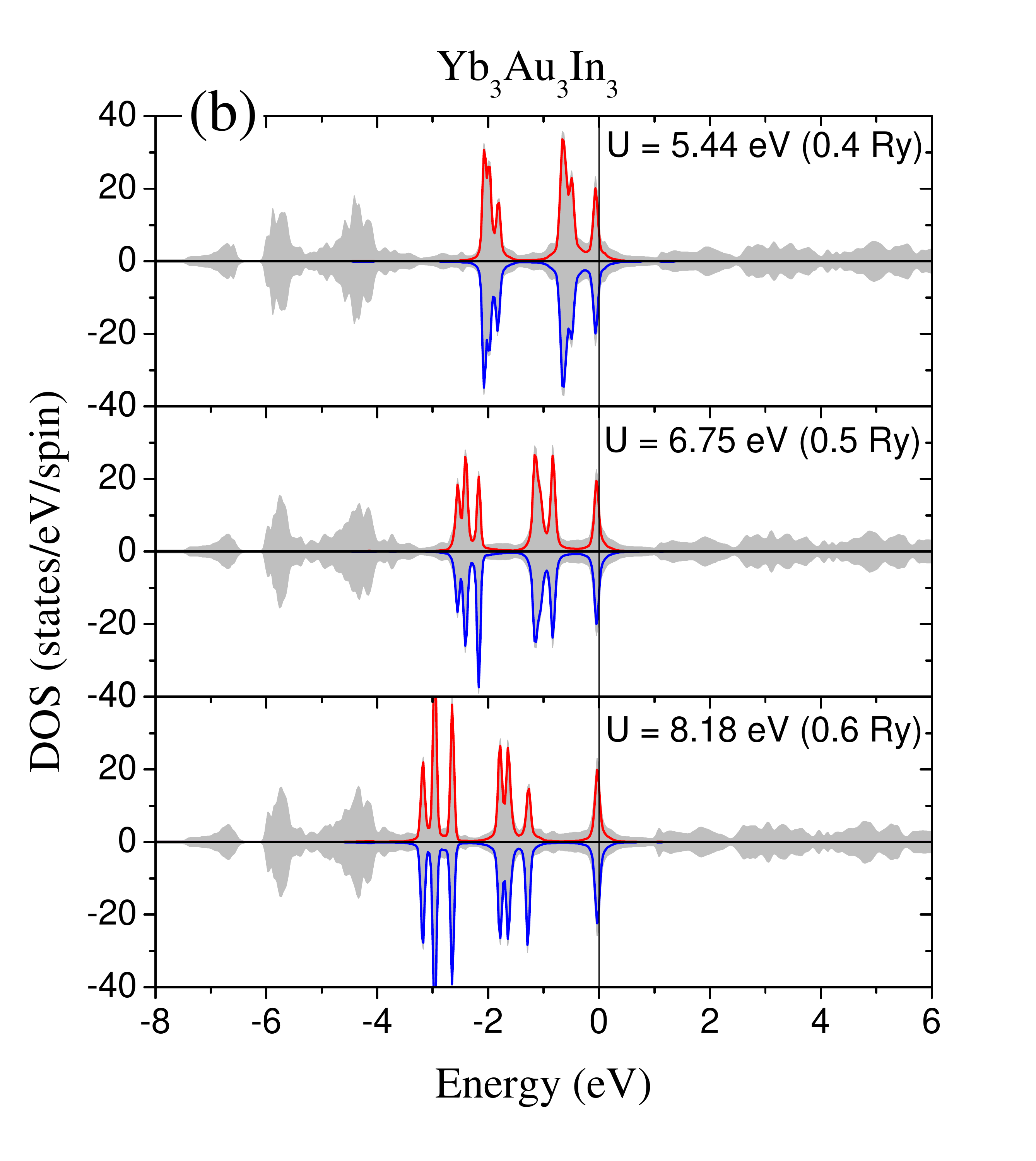}
\caption{(Color online) (a) Band structure of Yb${}_{3}$Au${}_{3}$In${}_{3}$ with the Au \textit{s} orbital character emphasized. The partially occupied 4\textit{f} level lies at E${}_{F}$, while the occupied levels are spread around -1.0 and -2.4 eV. (b) Total (shaded area) and partial density of states of YbAuIn associated with the Yb 4\textit{f }orbitals, calculated for different values of Hubbard \textit{U}. The partially occupied \textit{f} state is pinned at the Fermi level.}
\label{fig2}
\end{figure}

The calculated band structure of YbAuIn is illustrated in fig.~\ref{fig2}(a) and reveals the intermediate valent nature of YbAuIn, in agreement with magnetic susceptibility measurements~\cite{Zell1981, Chondroudi2011}. The flat bands located around 1.0 and 2.4 eV below the Fermi level ($E_F$) correspond to the occupied 4\textit{f} states of the Yb ion.  These states are split into 4\textit{f}$_{5/2}$ and 4\textit{f}$_{7/2 }$ manifolds under spin-orbit interaction of approximately 1.3 eV. Each of the two spin-orbit complexes are further split due to the anisotropy of the Coulomb interaction within the 4\textit{f} shell and possibly due to crystal field effect~\cite{Shick1999}. The additional flat band located at the $E_F$ corresponds to a partially occupied 4\textit{f} level. It is remarkable that the pinning of the 4\textit{f} state at the $E_F$ occurs even though, through manipulation of the density matrices~\cite{wu2008}, the calculations were initialized with the Yb 4\textit{f} shell fully occupied. Apparently, during the process of self-consistent relaxation one Yb 4\textit{f} band is split off from the rest of the 4\textit{f} complex, becomes partially occupied and is pinned to the $E_F$. This type of electronic structure is characteristic to intermediate valent heavy fermion compounds and it has been observed in other Yb-containing systems, such as YbBiPt~\cite{Oppeneer1998} and Yb$_4$As$_3$~\cite{Antonov1998}.

Figure~\ref{fig2}(b) illustrates the spin polarized density of states (DOS) of YbAuIn, projected on the Yb 4\textit{f} orbitals for different values of U. As the \textit{U}-parameter increases the partially occupied 4\textit{f} level remains pinned at the $E_F$ while the occupied levels shift down on the energy scale. The occupied and the partially occupied 4\textit{f}-states are well separated, indicating intermediate valent character for Yb in YbAuIn. To evaluate the valency of Yb the traces of the density matrices have been computed. The calculated 4\textit{f} occupations as a function of the \textit{U}-parameter indicate valences of 2.56, 2.60, and 2.66 for Yb for \textit{U} = 5.44, 6.75, and 8.16 eV respectively, somewhat larger that the experimental valence of $\sim$2.18~\cite{Zell1981, Chondroudi2011}. However, if we consider the fact that in the LAPW method the partial charges are considered only within the muffin-tin radii, and there is charge left in the interstitial region, the calculated valences are in reasonably good agreement with the experiment.

\begin{figure}
\centering
\includegraphics[scale=0.5]{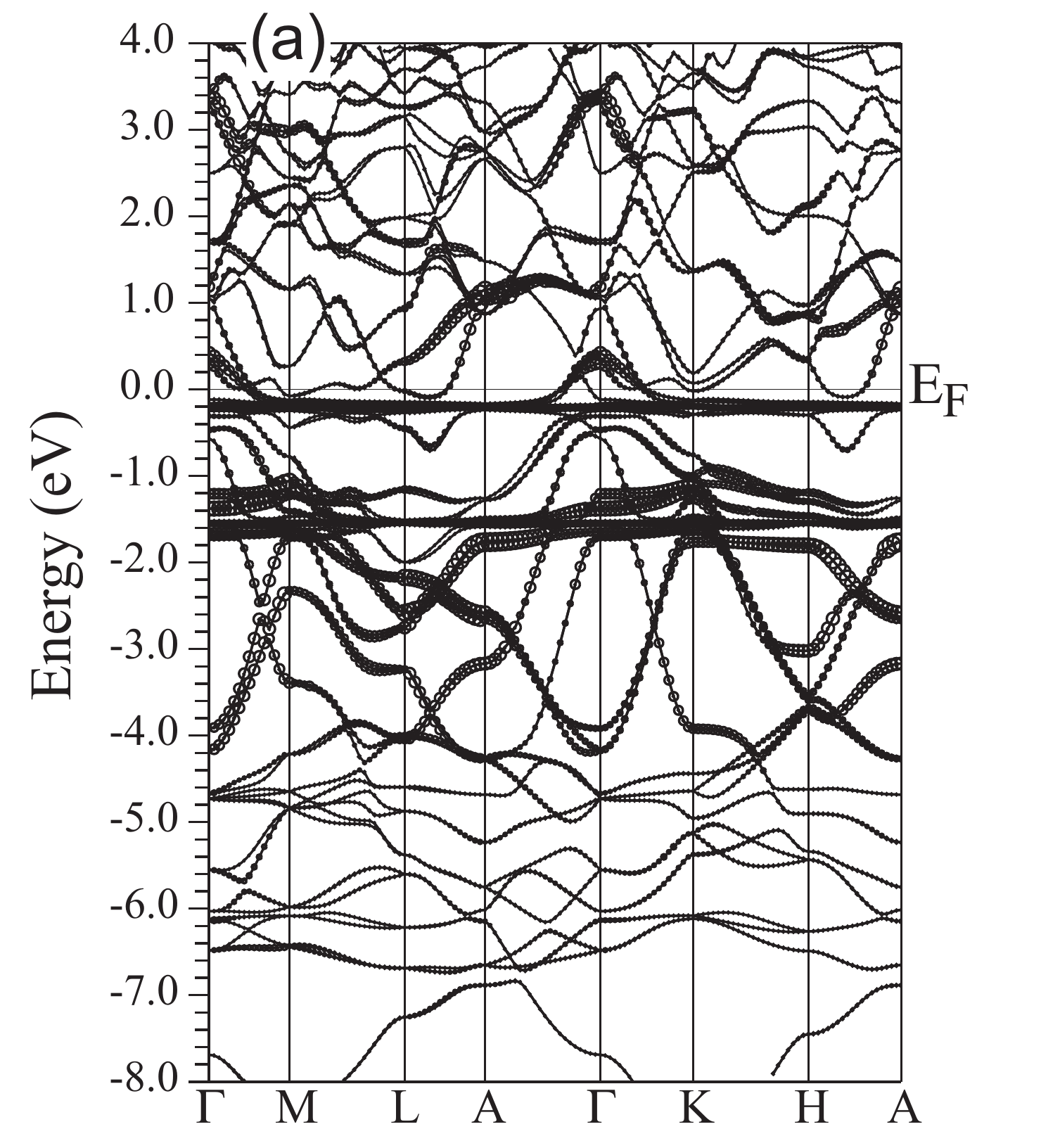}
\includegraphics[scale=0.28]{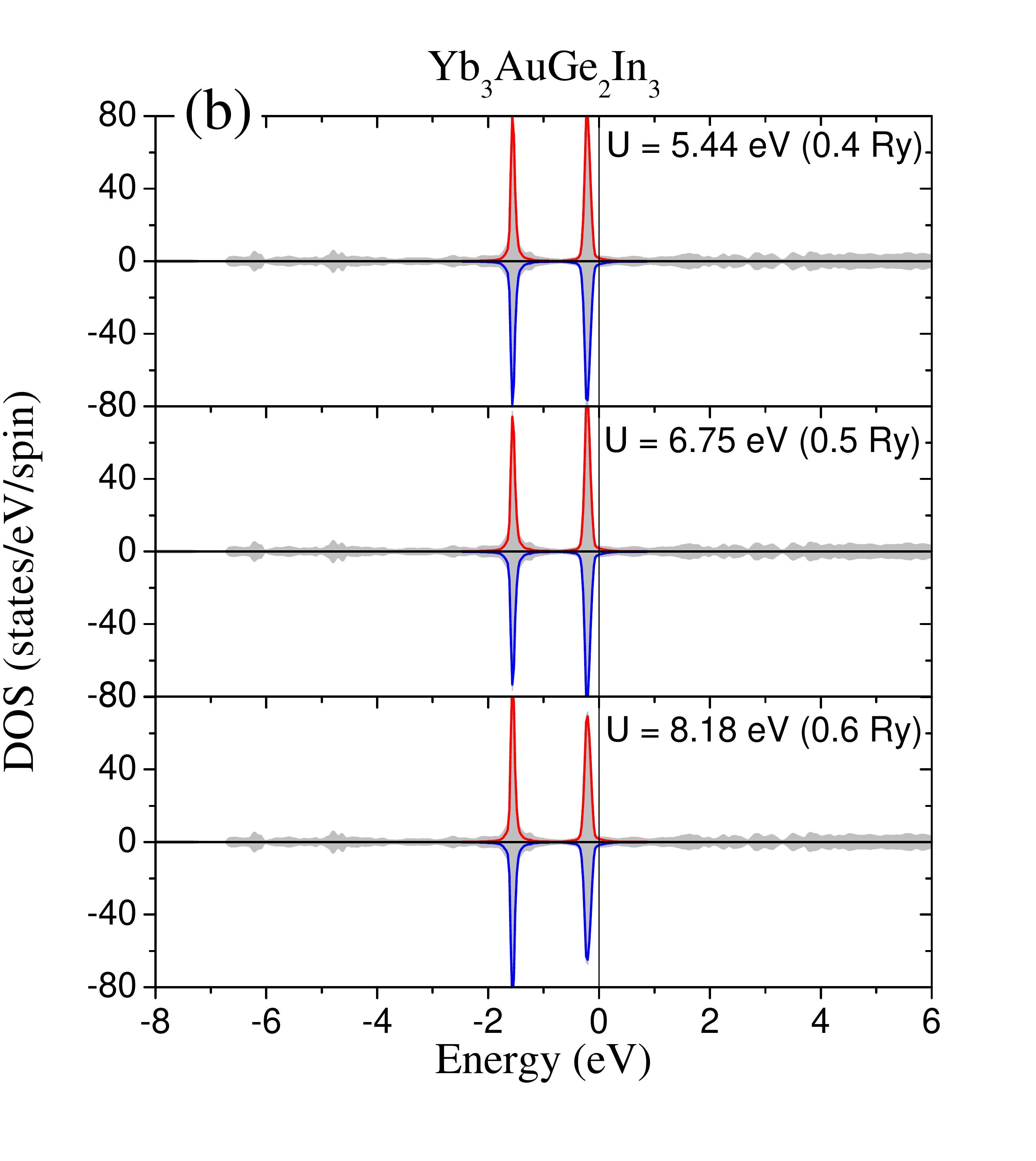}
\caption{(Color online) (a) Band structure of Yb${}_{3}$Au${}_{3}$In${}_{3}$ with the Au \textit{s} orbital character emphasized. The partially occupied 4\textit{f} level lies at E${}_{F}$, while the occupied levels are spread around -1.0 and -2.4 eV. (b) Total (shaded area) and partial density of states of YbAuIn associated with the Yb 4\textit{f }orbitals, calculated for different values of Hubbard \textit{U}. The partially occupied \textit{f} state is pinned at the Fermi level.}
\label{fig3}
\end{figure}

In the case of Yb$_{3}$AuGe$_{2}$In$_{3}$ the calculations indicate a quite different picture for the electronic structure and Yb valency. The band structure of Yb$_{3}$AuGe$_{2}$In$_{3}$, illustrated in fig.~\ref{fig3}(a) displays two flat bands located at approximately 0.2 and 1.5 eV below $E_F$.  These are the Yb 4\textit{f} bands, split into 4\textit{f}$_{5/2}$ and 4\textit{f}$_{7/2}$ manifolds, separated by the 1.3 eV spin orbit interaction. This can also be observed in fig.~\ref{fig3}(b), which illustrates the DOS associated with the Yb 4\textit{f} levels, calculated for different \textit{U}-parameters. It is apparent that the Yb 4\textit{f} states are occupied, located below $E_{F}$ and their position relative to $E_{F }$ does not change for different values of the Coulomb repulsion \textit{U}.

Because YbAuIn and Yb$_{3}$AuGe$_{2}$In$_{3}$ are isostructural and the latter is obtained by replacing two Au atoms in YbAuIn (or Yb$_{3}$Au$_{3}$In$_{3}$) by two Ge atoms, the differences in the electronic structure must be related to the presence of Ge. The physics of the Yb 4\textit{f} occupation can be understood by analyzing the positions of the 4\textit{f} levels relative to the Au 6\textit{s} bands in YbAuIn and Ge 4\textit{p} in Yb$_{3}$AuGe$_{2}$In$_{3}$, as illustrated with fat band representations in figs.~\ref{fig2}(a) and \ref{fig3}(a). In YbAuIn the Au 6\textit{s} states lie mostly below the Yb 4\textit{f}-level. When Au (5\textit{d}$^{10}$6\textit{s}$^{1}$) is replaced by Ge (3\textit{d}$^{10}$4\textit{s}$^{2}$4\textit{p}$^{2}$) the number of valence electrons in the system increases, therefore the Fermi level is shifted up in energy. Because the Ge 3\textit{p} band is not completely filled, the bands near E${}_{F }$(both below and above E${}_{F}$) display significant Ge \textit{p} character. Therefore, the upper Yb 4\textit{f}-level, which was initially pinned at the $E_{F}$, now lies below the top of Ge 3\textit{p} band, becomes fully occupied just below the $E_{F}$. Therefore, according to electronic structure calculations that include magnetic and relativistic effects as well as intra-atomic correlation, there is an essential difference between the two systems: in the YbAuIn the Yb ion has intermediate valent configuration, with one partially filled 4\textit{f} level and a valency of $\sim$2.6, while in Yb$_{3}$AuGe$_{2}$In$_{3}$, Yb is calculated to be closer to divalent state (Yb$^{2+}$).

The unusual behaviour of the lattice parameters, observed experimentally~\cite{Chondroudi2011}, can be associated with the valence configuration of the Yb.  Because Ge is smaller than Au, one would expect overall shorter lattice parameters forYb$_{3}$AuGe$_{2}$In$_{3}$ compared to those of YbAuIn. However, as shown above, in the presence of Ge, the valency of Yb becomes closer to Yb$^{2+}$ and because divalent Yb (Yb$^{2+}$) is larger than a mixed valent Yb (Yb$^{2.6+}$), the overall decrease in the lattice constants due to Ge is be counterbalanced by the increase in the size of Yb. If we consider the crystal structure of YbAuIn as being formed by alternating monoatomic sheets of In$_{3}$Au$_{2}$ and Yb$_{3}$Au stacked along the c-axis, we have the following scenario: on the one hand the area of the In$_{3}$Au$_{2}$ sheets shrinks when Au is replaced by Ge, but on the other hand, the separation between them increases, because the size of the Yb ion increases as the result of Yb$^{2.6+}$ $\rightarrow$ Yb${}^{2+}$ valence transition. The net effect is a decrease in the \textit{a}- and \textit{b}-parameters and an increase in the c-parameter of the hexagonal unit cell.

To verify this assumption, full structural relaxation have been carried out on two hypothetical compounds, LaAuIn and LuAuIn where La and Lu (La $>$ Lu) were chosen to simulate the divalent and intermediate valent Yb, respectively, and to avoid complications arising from the presence of localized \textit{f} electrons. The calculations were performed using VASP package~\cite{Kresse1996a, Kresse1996b} and included relaxation of internal parameters (atomic positions) and volume optimization. The calculated lattice parameters for LuAuIn are \textit{a} = 7.79 \AA, \textit{c} = 3.85 \AA and for LaAuIn are \textit{a} = 7.89 \AA, \textit{c} = 4.32 \AA, that is, when the smaller Lu is replaced by the larger La the \textit{a}-parameter increases only by $\sim$1.3\%, while the \textit{c}-parameter increases by $\sim$12.2\%. This indicates that the size of the ions that are located between the monoatomic sheets of In$_{3}$Au$_{2}$ primarily influences the length of the \textit{c}-parameter in the hexagonal unit cell, and has only a minor effect on the \textit{a}- and \textit{b}-parameters. This supports the premise that the unusual structural relationship between YbAuIn and Yb$_{3}$AuGe$_{2}$In$_{3}$, is associated with the Ge-induced Yb$^{2.6+}$ $\rightarrow$ Yb$^{2+}$ the valence transition.

In summary, using first-principles theoretical methods, the electronic structures of two isostructural Yb-intermetallics, YbAuIn and Yb$_{3}$AuGe$_{2}$In$_{3}$, have been investigated, focusing on the valence configuration of the Yb ion. YbAuIn is calculated as intermediate valence compound, with one partially filled Yb 4\textit{f} state, pinned to the Fermi level, which suggests that YbAuIn is a heavy fermion compound. Yb$_{3}$AuGe$_{2}$In$_{3}$ is calculated to be closer to integer valency with a completely occupied Yb 4\textit{f} shell. Furthermore, the electronic structure calculations combined with structural optimizations performed on LaAuIn and LuAuIn analogs, indicate that the longer \textit{c}-parameter in Yb$_{3}$AuGe$_{2}$In$_{3}$ compared with that in YbAuIn is attributable to the fact that the divalent Yb ion in Yb$_{3}$AuGe$_{2}$In$_{3}$ is larger than the intermediate valent Yb in YbAuIn. The change in the valence configuration is caused by the presence of Ge.

\begin{acknowledgements}

\noindent The computational work has been performed at NERSC, supported by the Office of Science of the US Department of Energy under Contract No. DE-AC02-05CH11231. I would like thank M. Chondroudi, M. G. Kanatzidis and S. D. Mahanti for valuable discussions regarding the electronic and magnetic properties of numerous Yb-containing intermetallics.

\end{acknowledgements}

\end{document}